\documentclass{SciPost}

\hypersetup{
    colorlinks,
    linkcolor={red!50!black},
    citecolor={blue!50!black},
    urlcolor={blue!80!black}
}

\usepackage[bitstream-charter]{mathdesign}
\DeclareSymbolFont{usualmathcal}{OMS}{cmsy}{m}{n}
\DeclareSymbolFontAlphabet{\mathcal}{usualmathcal}

\fancypagestyle{SPstyle}{
\fancyhf{}
\lhead{\colorbox{scipostblue}{\bf \color{white} ~SciPost Physics }}
\rhead{{\bf \color{scipostdeepblue} ~Submission }}

\fancyfoot[C]{\textbf{\thepage}}
}

\usepackage{xcolor}
\newcommand{\dd}{\ensuremath{\mathrm{d}}}
\newcommand{\br}{\ensuremath{\mathbf{r}}}

\begin{document}

\pagestyle{SPstyle}

\begin{center}{\Large \textbf{\color{scipostdeepblue}{
Dynamics of Polar-Core Spin Vortices in Inhomogeneous Spin-1 Bose-Einstein Condensates\\
}}}\end{center}

\begin{center}\textbf{
Zachary L. Stevens-Hough\textsuperscript{1},
Matthew J. Davis\textsuperscript{1} and
Lewis A. Williamson\textsuperscript{1$\star$}
}\end{center}

\begin{center}
{\bf 1} ARC Centre of Excellence for Engineered Quantum Systems, School of Mathematics and Physics, University of Queensland, St Lucia, Queensland 4072, Australia
\\[\baselineskip]
$\star$ \href{mailto:email1}{\small lewis.williamson@uq.edu.au}
\end{center}

\section*{\color{scipostdeepblue}{Abstract}}
\textbf{
In the easy-plane phase, a ferromagnetic spin-1 Bose-Einstein condensate is magnetized in a plane transverse to the applied Zeeman field. This phase supports polar-core spin vortices (PCVs), which consist of phase windings of transverse magnetization. Here we show that spin-changing collisions cause a PCV to accelerate down density gradients in an inhomogeneous condensate. The dynamics is well-described by a simplified model adapted from scalar systems, which predicts the dependence of the dynamics on trap tightness and quadratic Zeeman energy. In a harmonic trap, a PCV accelerates radially to the condensate boundary, in stark contrast to the azimuthal motion of vortices in a scalar condensate. In a trap that has a local potential maximum at the centre, the PCV exhibits oscillations around the trap centre, which persist for a remarkably long time. The oscillations coincide with the emission and reabsorption of axial spin waves, which reflect off the condensate boundary.
}

\vspace{\baselineskip}

\noindent\textcolor{white!90!black}{%
\fbox{\parbox{0.975\linewidth}{%
\textcolor{white!40!black}{\begin{tabular}{lr}%
  \begin{minipage}{0.6\textwidth}%
    {\small Copyright attribution to authors. \newline
    This work is a submission to SciPost Physics. \newline
    License information to appear upon publication. \newline
    Publication information to appear upon publication.}
  \end{minipage} & \begin{minipage}{0.4\textwidth}
    {\small Received Date \newline Accepted Date \newline Published Date}%
  \end{minipage}
\end{tabular}}
}}
}


\vspace{10pt}
\noindent\rule{\textwidth}{1pt}
\tableofcontents
\noindent\rule{\textwidth}{1pt}
\vspace{10pt}

\section{Introduction}

Topological defects underpin a variety of physical processes, including ordering dynamics~\cite{Bray1994}, quantum turbulence~\cite{vinen2002} and topological phase transitions~\cite{kosterlitz2017}. The interplay between density inhomogeneity and topological defects provides a rich area of exploration, giving rise to effects such as defect pinning~\cite{nelson1993,pazo2004,campbell2014,stockdale2021,anderson1975,warszawski2012}, modified critical dynamics~\cite{delcampo2010,yi2020,kim2022,rabga2023} and novel phases of matter~\cite{gallemi2020,roccuzzo2020,klaus2022}. Spin-1 Bose-Einstein condensates support multiple spin phases and associated topological excitations~\cite{ueda2014}, are well isolated from the environment, and can be manipulated with high precision.  Various topological defects and textures have been observed experimentally in this system~\cite{Leanhardt2003a,Sadler2006a,Choi2012a,ray2014,Seo2015a,weiss2019}, and non-destructive imaging techniques~\cite{Higbie2005a} have enabled \emph{in situ} studies of vortex interactions~\cite{Seo2016a,Kang2019a} and the role of vortices in the early stages of phase ordering~\cite{Sadler2006a}.

In the easy-plane phase a ferromagnetic spin-1 condensate is magnetized in the transverse plane with $\mathrm{SO}(2)$ spin symmetry~\cite{Murata2007a}. In two dimensions circulation of transverse spin, with no additional global phase circulation, gives rise to polar-core spin vortices (PCVs)~\cite{isoshima2001,Sadler2006a}. These play a fundamental role in both equilibrium~\cite{Mukerjee2006a,kobayashi2019,underwood2022} and non-equilibrium~\cite{Sadler2006a,Saito2007b,Saito2007a,Uhlmann2007a,Lamacraft2007a,Kudo2015a,williamson2016a,williamson2016b,Schmied2019a} properties of the system. A point-vortex model of PCVs shows that opposite (same) charged PCVs tend to attract (repel), which is attributable to spin-changing collisions in the vortex core~\cite{Turner2009,Williamson2016c,williamson2021}. To date, theoretical studies of PCV dynamics have focused on homogeneous spinor condensates. On the other hand, spinor condensates realised in experiments so far have inhomogeneous densities~\cite{Sadler2006a,prufer2018,prufer2022,huh2020,huh2023}.
It is well known that a vortex in a scalar condensate is sensitive to the background condensate profile induced by a trapping potential, which causes vortex procession in an axisymmetric trap~\cite{anderson2000,freilich2010,rokhsar1997,jackson1999,fedichev1999,svidzinsky2000,svidzinsky2000b,lundh2000,tempere2000,mcgee2001,sheehy2004,fetter2001}.
The corresponding question in regards to PCVs --- how does a single PCV move in an inhomogeneous density profile --- has, to our knowledge, not been explored.

In this work we show how PCVs move in inhomogeneous density profiles, focusing on axisymmetric traps.
We show that a PCV tends to move down density gradients, irrespective of the sign of the PCV charge.
Hence in a harmonic trap a PCV moves radially outward, in stark contrast to the azimuthal motion of a vortex in a scalar condensate. This radial motion is associated with a separation (``stretching'') of the component circulations that make up a PCV, which tend to move azimuthally in the background field but are bound by the spin exchange energy. A trap with a local potential maximum at the centre produces a local minimum in condensate density. An off-center PCV then moves radially inward. Rather than settling at the trap centre, we find that the PCV instead undergoes periodic perturbations in displacement, coinciding with the emission and reabsorption of axial spin-wave excitations that reflect off the condensate boundary. These dynamics are remarkably long-lived, decaying as a slow power law in time for long times. We analyse the dynamics of incompressible and compressible energy during the PCV dynamics. We find that the incompressible kinetic energy decreases as the vortex moves to lower densities. In the harmonic trap this energy is primarily converted into internal spin energy. In the trap with a local potential maximum at the centre the energy is converted back-and-forth between compressible and incompressible kinetic energy, consistent with the emission and reabsorption of axial spin waves.

This paper is organised as follows. In Sec.~\ref{sec:PCVs} we introduce the system and PCVs. In Sec.~\ref{sec:model} we present a model of PCV dynamics in an inhomogeneous background. We then present numerical results of PCV dynamics in a harmonic trap (Sec.~\ref{sec:harmonic}) and in a box trap with a local potential maximum at the centre (Sec.~\ref{sec:osc}), comparing with the model in Sec.~\ref{sec:model}. In Sec.~\ref{sec:energy} we analyse the energy exchanges that occur during the dynamics. We conclude in Sec.~\ref{sec:conclusion}.

\section{Polar-core spin vortices}\label{sec:PCVs}

A spin-1 condensate can be described by a three component spinor field $\Psi=(\psi_1,\psi_0,\psi_{-1})^T$, with $\psi_m$ describing atoms in the magnetic sublevel $m\in \{-1,0,1\}$. In a quasi-2D geometry in the mean-field approximation, the system is described by the Hamiltonian~\cite{Kawaguchi2012R}
\begin{equation}\label{eq:Hamiltonian}
H = \int\dd^2\br\, \left\{
\sum_{m=-1}^{1} \psi_m^*(\br)\left[-\frac{\hbar^2}{2M}\nabla^2+U(\br)+qm^2\right]\psi_m(\br)
+ \frac{g_n}2n(\br)^2 + \frac{g_s}2|\mathbf F(\br)|^2
\right\}.
\end{equation}
Here $n(\br) = \sum_m \psi_m^*\psi_m=\Psi^\dag\Psi$ is the areal number density and $\mathbf F(\br)=\sum_{\mu=x,y,z}\Psi^\dag f_\mu\Psi\, \hat{\mathbf s}_\mu$ is the areal spin density with $f_\mu$ the spin-1 Pauli matrices in the direction $\hat{\mathbf s}_\mu$. Explicitly, the three components of $\mathbf{F}$ are
\begin{equation}
\begin{split}
F_x=&\sqrt{2}\operatorname{Re}\left(\psi_1^*\psi_0+\psi_0^*\psi_{-1}\right),\\
F_y=&\sqrt{2}\operatorname{Im}\left(\psi_1^*\psi_0+\psi_0^*\psi_{-1}\right),\\
F_z=&|\psi_1|^2-|\psi_{-1}|^2.
\end{split}
\end{equation}
The trapping potential $U(\mathbf{r})$ is in addition to any trap confining the condensate to a 2D geometry and results in an inhomogeneous density profile in the 2D plane. In this work we will restrict our analysis to axisymmetric traps $U(\mathbf{r})=U(r)$. The quadratic Zeeman energy $q$ controls the single-particle level spacings along the quantization axis $\hat{\mathbf s}_z$, and can be tuned in experiments using a variety of techniques~\cite{StamperKurn2013a}. Any linear Zeeman shift has been removed by transforming $\Psi$ into a frame rotating at the Larmor frequency. The spin and density interaction strengths are $g_s$ and $g_n$ respectively. We consider ferromagnetic spin interactions, $g_s<0$, as realised in $^{87}$Rb~\cite{Schmaljohann2004,Chang2004a} and $^7$Li~\cite{huh2020} condensates.

The time-evolution of each spin level is governed by the spin-1 Gross-Pitaevskii equations~\cite{Kawaguchi2012R}
\begin{equation}
i\hbar\frac{\partial \psi_m}{\partial t}=\left[-\frac{\hbar^2}{2M}\nabla^2+U(\br)+qm^2+g_n n\right]\psi_m+g_s\sum_{m'=-1}^1\mathbf F\cdot f_{mm'}\psi_{m'}.\label{eq:time_evolution_psi}
\end{equation}
We take $g_n/|g_s|=10$ so that the dynamics are predominantly confined to the spin degrees of freedom. We expect this regime to be representative of the behaviour of an $^{87}$Rb condensate, which has $|g_s|\ll g_n$~\cite{kempen2002,widera2006}. The spin time $t_s\equiv\hbar/q_0$ and the spin healing length $\xi_s\equiv \hbar/\sqrt{Mq_0}$ are then convenient time and length scales. Here $q_0=2|g_s|n_0$ with $n_0$ the condensate density at the trap minimum.

The ratio of $q$ to spin interaction energy results in a rich variety of magnetic ground states~\cite{Ohmi1998a,Stenger1998a}. In a uniform system, $q=q_0$ is a quantum critical point separating the unmagnetized (``polar'') phase ($q>q_0$) from the magnetized phases ($q<q_0$)~\cite{Stenger1998a}. When $0<q<q_0$, the ground state of Eq.~\eqref{eq:Hamiltonian} is magnetized in a plane perpendicular to the external Zeeman field, termed the easy-plane phase.
The ground state manifold consists of both a $\mathrm U(1)$ global gauge symmetry and an $\mathrm{SO}(2)$ spin symmetry $e^{-if_z\varphi}$~\cite{Murata2007a}.
A $2\pi\kappa\, (\kappa=\pm 1)$ phase winding of $\varphi$ gives rise to circulations $-2\pi m\kappa$ in the spin components $\psi_m$, and a $2\pi\kappa$ circulation of transverse spin $F_\perp=F_x+iF_y$.
With no additional global phase winding, this circulation can be identified as a singly-charged polar-core spin vortex, with vortex core filled by the (circulation-free) $m=0$ component~\cite{isoshima2001,Sadler2006a},
\begin{equation}\label{pcvstate}
    \Psi(\mathbf{r})=\sqrt{\frac{n}{2}}\left(\begin{array}{c}\sin\beta g_1(\mathbf{r})e^{-i\kappa\phi(\mathbf{r})}\\\sqrt{2}\cos\beta g_0(\mathbf{r})\\\sin\beta g_{-1}(\mathbf{r})e^{i\kappa\phi(\mathbf{r})}\end{array}\right).
\end{equation}
Here $\phi(\mathbf{r})=\operatorname{phase}(z)$ is the phase of the complex number $z=x+iy$, $\cos(2\beta)=q/q_0$, and $g_m(\mathbf{r})\ge 0$ accounts for the density of the vortex core~\cite{Lovegrove2012a,Lovegrove2016a} with $g_1(\mathbf{r})=g_{-1}(\mathbf{r})$, $g_{\pm 1}(\mathbf{0})=0$ and $g_m(\mathbf{r})\rightarrow 1$ for $r\gg\xi_s$. Here and below we take $\psi_0$ to be real without loss of generality. A notable feature of Eq.~\eqref{pcvstate} is the equal and opposite phases of the $\psi_{\pm 1}$ components. This ensures that the spin exchange component $E_\mathrm{se}$ of the spin interaction energy is minimised, with
\begin{equation}\label{Ese}
    E_\mathrm{se}=g_s\int\dd^2\br\,\left[\psi_0^2\psi_1^*\psi_{-1}^*+(\psi_0^*)^2\psi_1\psi_{-1}\right].
\end{equation}
The spin interaction energy is responsible for spin-changing collisions $0+0\leftrightarrow 1+(-1)$ and is absent in multi-component condensates of distinct species or distinct hyperfine manifolds.

In an inhomogeneously trapped system the ground state is easy-plane for $q\lesssim 2|g_s|n(\mathbf{r})$ and hence the phase depends on the inhomogeneous condensate density. In this work we consider axisymmetric traps with a single phase boundary at $r=r_{F_\perp}$, such that the system is easy-plane for $r<r_{F_\perp}$ and polar for $r>r_{F_\perp}$. Approximating $n(\mathbf{r})$ by the Thomas-Fermi density~\cite{baym1996},
\begin{equation}\label{nTF}
    n_\mathrm{TF}(\mathbf{r})=n_0\left(1-\frac{U(\mathbf{r})}{\mu_\mathrm{TF}}\right),
\end{equation}
we define $r_{F_\perp}$ via $U(r_{F_\perp})=\mu_\mathrm{TF}(1-q/q_0)$, with $\mu_\mathrm{TF}=g_n n_0$ the approximate chemical potential.

\section{A model of PCV dynamics in an inhomogeneous condensate}\label{sec:model}
\subsection{Equations of motion for PCV dynamics}
To gain an intuitive understanding of how a singly-charged PCV moves in an inhomogeneous density, we formulate a model adapted from scalar systems~\cite{nilsen2006,jezek2008,santos2016,groszek2018}. A PCV consists of equal and oppositely charged vortices in the $m=\pm 1$ components, see Eq.~\eqref{pcvstate}. Following~\cite{groszek2018}, we write the wavefunctions for these vortex fields as
\begin{equation}\label{psimv}
\begin{split}
\psi_1(\br,t)=&(z^*(\br)-z_1(t))A_1(\mathbf{r},t)e^{i\theta_1(\mathbf{r},t)},\\
\psi_{-1}(\br,t)=&(z(\br)-z_{-1}(t))A_{-1}(\mathbf{r},t)e^{i\theta_{-1}(\mathbf{r},t)},
\end{split}
\end{equation}
with $z_m(t)=x_m(t)-i\kappa m y_m(t)$ and $z(\br)=x+i\kappa y$. The complex factors $z^*(\mathbf{r})-z_1$ and $z(\mathbf{r})-z_{-1}$ account for both a circular vortex phase circulation (charge $-m\kappa$) and the density dip at the vortex core. The centre of circulation for the vortex in component $\psi_m$ is at $\mathbf{x}_m=(x_m,y_m)$. The background amplitude $A_m$ and phase profile $\theta_m$ describe the remaining portion of $\psi_m$ and account for inhomogeneities caused by the trap~\cite{fetter2009}. Additionally, and unique to the spinor system, $A_m$ and $\theta_m$ may be affected by separation of the $\psi_{\pm 1}$ vortices due to spin-exchange interactions~\cite{williamson2021,takeuchi2021}.

An equation of motion for $\mathbf{x}_m$ can be obtained by evolving the state~\eqref{psimv} using Eq.~\eqref{eq:time_evolution_psi} and evaluating the resulting expressions at $\mathbf{r}=\mathbf{x}_m$~\cite{groszek2018}. This gives
\begin{equation}\label{vm1}
-i\hbar \dot{z}_m(t)A_m(\mathbf{x}_m)e^{i\theta_m(\mathbf{x}_m)}=-\frac{\hbar^2}{M}e^{i\theta_m(\mathbf{x}_m)} \left(\nabla A_m+iA_m\nabla\theta_m\right)\big|_{\mathbf{x}_m}\cdot \mathbf{u}_m+g_s\psi_0(\mathbf{x}_m)^2\psi_{-m}^*(\mathbf{x}_m),
\end{equation}
with $\mathbf{u}_1=\nabla z^*(\mathbf{r})=(1,-\kappa i)$, $\mathbf{u}_{-1}=\nabla z(\mathbf{r})=(1,\kappa i)$, and we have used that
\begin{equation}\label{gradexpand}
\nabla^2\psi_m\big|_{\mathbf{x}_m}=2\mathbf{u}_m\cdot\nabla \left(A_me^{i\theta_m}\right)\big|_{\mathrm{x}_m}.
\end{equation}
Terms from Eq.~\eqref{eq:time_evolution_psi} containing a factor of $\psi_m$ are absent from Eq.~\eqref{vm1} since $\psi_m(\mathbf{x}_m)=0$. Rearranging Eq.~\eqref{vm1} for $\dot{z}_m$ and writing in vector format by equating real and imaginary parts gives,
\begin{equation}\label{vm}
\dot{\mathbf{x}}_m=\frac{\hbar}{M}\left(\kappa m\hat{\mathbf{z}}\times\nabla \ln A_m+\nabla \theta_m\right)\big|_{\mathbf{x}_m}+\mathbf{J}_m,
\end{equation}
where 
\begin{equation}\label{source}
\mathbf{J}_m=\frac{|g_s|}{\hbar}\left(\operatorname{Im}\frac{\psi_0(\mathbf{x}_m)^2\psi_{-m}^*(\mathbf{x}_m)}{A_m(\mathbf{x}_m)e^{i\theta_m(\mathbf{x}_m)}},\kappa m\operatorname{Re}\frac{\psi_0(\mathbf{x}_m)^2\psi_{-m}^*(\mathbf{x}_m)}{A_m(\mathbf{x}_m)e^{i\theta_m(\mathbf{x}_m)}}\right).
\end{equation}
The first term on the right-hand side of Eq.~\eqref{vm} arises from the kinetic energy in Eq.~\eqref{eq:time_evolution_psi} and has the same form as for a vortex in a scalar condensate. In an axisymmetric trap with no imposed background flow, this term induces azimuthal vortex motion~\cite{groszek2018}. The second term $\mathbf{J}_m$ is a source term arising from the spin exchange component of the spin interaction energy, Eq.~\eqref{Ese}, and is the key feature that will distinguish the PCV dynamics. In the absence of $\mathbf{J}_m$ and any background superfluid flow, the $m=\pm 1$ vortices will circulate the trap centre in opposite directions, causing a separation of the $m=\pm 1$ vortex cores. In the presence of $\mathbf{J}_m$ this ``stretching'' cannot occur indefinitely, as the components are bound together by the spin exchange energy~\cite{Turner2009,Williamson2016c}. Note the PCV stretching deforms the state Eq.~\eqref{pcvstate}, as the $m=\pm 1$ vortex cores no longer coincide. Also, despite the separation of the $m=\pm 1$ vortex cores, the total mass current remains zero since the total $m=\pm 1$ currents cancel~\cite{Turner2009}.

At this point it is useful to make some general statements about the dynamics by considering the symmetries of the problem. The Hamiltonian is symmetric under inversion of $F_z$, which transforms $m\rightarrow -m$. For an axisymmetric trap the Hamiltonian is also symmetric under a reflection across any radial line, in particular one through the point $\mathbf{x}_v=(\mathbf{x}_1+\mathbf{x}_{-1})/2$. The PCV position $\mathbf{x}_v$ coincides with the centre of phase winding of $F_\perp$. We assume an initial state of the form Eq.~\eqref{pcvstate} but with the initial PCV position $\mathbf{x}_v=\mathbf{x}_1=\mathbf{x}_{-1}$ displaced from the origin and positioned on the $x$-axis. Due to the symmetry of the Hamiltonian and the initial state, the time evolution evolves the fields $\psi_1(\mathbf{r})$ and $\psi_{-1}(\mathbf{r})$ identically aside from a reflection in the $x$-axis, hence
\begin{equation}\label{symmetry}
\psi_1(x,y)=\psi_{-1}(x,-y)
\end{equation}
throughout the time evolution. This gives $(x_1,y_1)=(x_{-1},-y_{-1})$. Hence, for an axisymmetric trap, the PCV position $\mathbf{x}_v$ is confined to move radially, whereas the ``stretch coordinate'' $\mathbf{s}=\mathbf{x}_1-\mathbf{x}_{-1}$ is orthogonal to this.

We now write Eq.~\eqref{vm} in terms of $\mathbf{x}_v$ and $\mathbf{s}$. Evaluating $\nabla\psi_m\big|_{\mathbf{x}_m}$ and $\nabla^2\psi_m\big|_{\mathbf{x}_m}$ from Eq.~\eqref{psimv} and using the symmetry constraint Eq.~\eqref{symmetry} gives
\begin{equation}\label{symmetry2}
\begin{split}
A_1(\mathbf{x}_1)e^{i\theta_1(\mathbf{x}_1)}&=A_{-1}(\mathbf{x}_{-1})e^{i\theta_{-1}(\mathbf{x}_{-1})},\\
\mathbf{u}_1\cdot\nabla\left(A_1e^{i\theta_1}\right)\big|_{\mathbf{x}_1}&=\mathbf{u}_{-1}\cdot\nabla\left(A_{-1}e^{i\theta_{-1}}\right)\big|_{\mathbf{x}_{-1}},
\end{split}
\end{equation}
which can be combined to give
\begin{equation}\label{symmetry3}
\frac{\partial X_1}{\partial x}\Bigg|_{\mathbf{x}_1}=\frac{\partial X_{-1}}{\partial x}\Bigg|_{\mathbf{x}_{-1}},\hspace{1cm}\frac{\partial X_1}{\partial y}\Bigg|_{\mathbf{x}_1}=-\frac{\partial X_{-1}}{\partial y}\Bigg|_{\mathbf{x}_{-1}},\hspace{1cm}\left(X=\ln A,\,\theta\right).
\end{equation}
Equation~\eqref{vm} and Eq.~\eqref{symmetry3} give the equations of motion
\begin{equation}\label{eqmotion1}
\begin{split}
\dot{\mathbf{s}}&=\frac{2\hbar}{M}\left(\kappa\frac{\partial\ln A_1}{\partial x}+\frac{\partial\theta_1}{\partial y}\right)\Bigg |_{\mathbf{x}_1}\hat{\mathbf{y}}+\sum_m m\mathbf{J}_m,\\
\dot{\mathbf{x}}_v &=\frac{\hbar}{M}\left(-\kappa\frac{\partial\ln A_1}{\partial y}+\frac{\partial\theta_1}{\partial x}\right)\Bigg |_{\mathbf{x}_1}\hat{\mathbf{x}}+\frac{1}{2}\sum_m \mathbf{J}_m.
\end{split}
\end{equation}
Equations~\eqref{eqmotion1} can equivalently be written in terms of the background fields $A_{-1}$ and $\theta_{-1}$ using Eq.~\eqref{symmetry3}.

\subsection{Small-stretching approximation}
Further progress can be made by assuming the stretch coordinate is small. The source $\mathbf{J}_m$ depends on the stretch coordinate $\mathbf{s}$ via $\psi_{-m}^*(\mathbf{x}_m)=\psi_{-m}^*(\mathbf{x}_{-m}+m\mathbf{s})$. Expanding this to lowest order in $\mathbf{s}$ and using Eq.~\eqref{psimv} gives,
\begin{equation}\label{psirelation}
\begin{split}
\psi_{-m}^*(\mathbf{x}_m)&\approx m\mathbf{s}\cdot\nabla\psi_{-m}^*\big|_{\mathbf{x}_{-m}}\\
&=mA_{-m}(\mathbf{x}_{-m})e^{-i\theta_{-m}(\mathbf{x}_{-m})}\mathbf{s}\cdot\mathbf{u}_{-m}^*\\
&=mA_m(\mathbf{x}_m)e^{-i\theta_{-m}(\mathbf{x}_{-m})}\mathbf{s}\cdot\mathbf{u}_m.
\end{split}
\end{equation}
In the last line we have used $\mathbf{u}_{-m}^*=\mathbf{u}_m$ and Eq.~\eqref{symmetry2}, which gives $A_1(\mathbf{x}_1)=A_{-1}(\mathbf{x}_{-1})$. Substituting Eq.~\eqref{psirelation} into Eq.~\eqref{source} gives a source term that depends on the stretching as well as the background fields $\psi_0(\mathbf{x}_m)$ and $e^{-i(\theta_m(\mathbf{x}_m)+\theta_{-m}(\mathbf{x}_{-m}))}$. We approximate these background fields by their $\mathbf{s}=0$ values, which is consistent with the linear expansion in Eq.~\eqref{psirelation}, $\theta_m(\mathbf{x}_m)+\theta_{-m}(\mathbf{x}_{-m})\approx 0$ (which minimises the the spin exchange energy Eq.~\eqref{Ese} for real $\psi_0$) and $\psi_0(\mathbf{x}_m)\approx \psi_0(\mathbf{x}_v)$. This gives
\begin{equation}\label{Jmapprox}
    \mathbf{J}_m\approx \frac{\kappa |g_s| \psi_0(\mathbf{x}_v)^2}{\hbar}\hat{\mathbf{z}}\times\mathbf{s}.
\end{equation}

We assume the background fields $A_m$ and $\theta_m$ vary slowly with $\mathbf{s}$ and hence approximate their gradients in Eq.~\eqref{eqmotion1} by their $\mathbf{s}=0$ values obtained from $\psi_1(\mathbf{r})=\psi_{-1}^*(\mathbf{r})$. Consistency with Eq.~\eqref{symmetry3} then requires $\partial\ln A_1/\partial y\big|_{\mathbf{x}_1}=0$ and $\partial\theta_1/\partial x\big|_{\mathbf{x}_1}=0$. Substituting Eq.~\eqref{Jmapprox} into Eq.~\eqref{eqmotion1} then gives the equations of motion
\begin{align}
\label{sapprox}
\dot{\mathbf{s}}&\approx \frac{2\hbar}{M}\left(\kappa\hat{\mathbf{z}}\times\nabla \ln A_1+\nabla \theta_1\right)\big|_{\mathbf{x}_1},\\
\label{fapprox1}
\dot{\mathbf{x}}_v&\approx \frac{\kappa |g_s|\psi_0(\mathbf{x}_v)^2}{\hbar}\hat{\mathbf{z}}\times\mathbf{s}.
\end{align}
Equation~\eqref{sapprox} shows that density and phase gradients will stretch a PCV and Eq.~\eqref{fapprox1} shows that this stretching will cause a PCV to move in a direction orthogonal to the stretching.

We also briefly remark that interactions with other PCVs could be included in Eq.~\eqref{sapprox} by including in $\nabla\theta_1$ the flow field of the additional PCVs. In a homogeneous system this predicts $a\approx \psi_0(\mathbf{x}_v)^2/n_0$, with $a$ introduced empirically in~\cite{Williamson2016c,williamson2021} and related to the ``spring constant'' of the stretch energy. Using a uniform background $\psi_0(\mathbf{x}_v)^2/n_0=\cos\beta$ gives $a=(1+q/q_0)/2$, which correctly predicts an increase in $a$ with $q/q_0$ but overestimates $a$ for $q/q_0\lesssim 0.5$~\cite{williamson2021}.

\subsection{Qualitative description of PCV dynamics}
Equation~\eqref{sapprox} has limited utility as a direct computational tool, since the calculation of $\nabla \theta_1$ is not straightforward~\cite{groszek2018}. In addition to this, the dynamics of the stretch coordinate is susceptible to intrinsic damping~\cite{williamson2021}. Ignoring interactions with the other spin components, an approximate analytic calculation for $\theta_1$ gives~\cite{sheehy2004,jezek2008,groszek2018}
\begin{equation}\label{theta1approx}
    \nabla\theta_1\big|_{\mathbf{x}_1}\sim -\kappa\left(\ln \left|\xi_R\nabla\ln A_1\right|\right)\hat{\mathbf{z}}\times\nabla\ln A_1\big |_{\mathbf{x}_1},
\end{equation}
where $\xi_R\sim \xi_s$\footnote{In~\cite{sheehy2004,jezek2008} $\xi_R$ is replaced by $|\mathbf{r}-\mathbf{x}_1|$, which results in $\nabla\theta_1\big|_{\mathbf{x}_1}$ diverging. In~\cite{groszek2018} a regularization of this divergence was found to better match numerical results.}. Hence, crudely, $\dot{\mathbf{s}}\sim (\hbar\kappa/M)\hat{\mathbf{z}}\times \nabla \ln A_1\big|_{\mathbf{x}_1}$, ignoring variation with $\mathbf{x}_1$ due to the logarithmic term $\ln|\xi_R\nabla\ln A_1|\big|_{\mathbf{x}_1}$ in Eq.~\eqref{theta1approx}. We set
\begin{equation}\label{A1TF}
A_1\approx \sqrt{\frac{n_\mathrm{TF}(\mathbf{r})}{4}\left(1-\frac{q}{2|g_s|n_\mathrm{TF}(\mathbf{r})}\right)}
\end{equation}
and
\begin{equation}\label{psi0TF}
    \psi_0(\mathbf{r})\approx \sqrt{\frac{n_\mathrm{TF}(\mathbf{r})}{2}\left(1+\frac{q}{2|g_s|n_\mathrm{TF}(\mathbf{r})}\right)},
\end{equation}
which follow from Eq.~\eqref{pcvstate} with $n$ replaced by $n_\mathrm{TF}$ and $\cos(2\beta)$ replaced by $q/(2|g_s|n_\mathrm{TF})$. Using Eq.~\eqref{A1TF} gives a qualitative estimate for $\dot{\mathbf{s}}$, which can be substituted into the time derivative of Eq.~\eqref{fapprox1} to obtain a qualitative estimate for $\ddot{\mathbf{x}}_v$,
\begin{equation}\label{fapprox2}
\begin{split}
\ddot{\mathbf{x}}_v &\approx \frac{\kappa |g_s|\psi_0(\mathbf{x}_v)^2}{\hbar}\hat{\mathbf{z}}\times\dot{\mathbf{s}}\\
&\sim \frac{\xi_s^2}{\mu_\mathrm{TF} t_s^2}\left(1+\frac{q}{2|g_s|n_\mathrm{TF}(\mathbf{x}_v)}\right)\left(1-\frac{q}{2|g_s|n_\mathrm{TF}(\mathbf{x}_v)}\right)^{-1}\nabla U\big|_{\mathbf{x}_v}.
\end{split}
\end{equation}
We have neglected the term proportional to $\dot{\psi}_0(\mathbf{x}_v)$ in the first line of Eq.~\eqref{fapprox2} since it is proportional to $\mathbf{s}$ and hence small. 

The key insight provided by the second line in Eq.~\eqref{fapprox2} is that a PCV will tend to accelerate in a direction of increasing potential, independent of the sign of the PCV charge, with a magnitude proportional to $|\nabla U|$.
This is in stark contrast to the azimuthal motion of a scalar vortex.
Equation~\eqref{fapprox2} also predicts the acceleration will increase with $q$ due to an increase in $\psi_0^2/A_1^2$. Note that $|\nabla U|/\mu_\mathrm{TF}\sim 1/r_{F_\perp}$, hence the PCV will move a distance $r_{F_\perp}$ over a time scale $t\sim (r_{F_\perp}/\xi_s)t_s$.

\begin{figure}[!t]
  \includegraphics[trim=0cm 8cm 0cm 8cm,clip=true,width=\columnwidth]{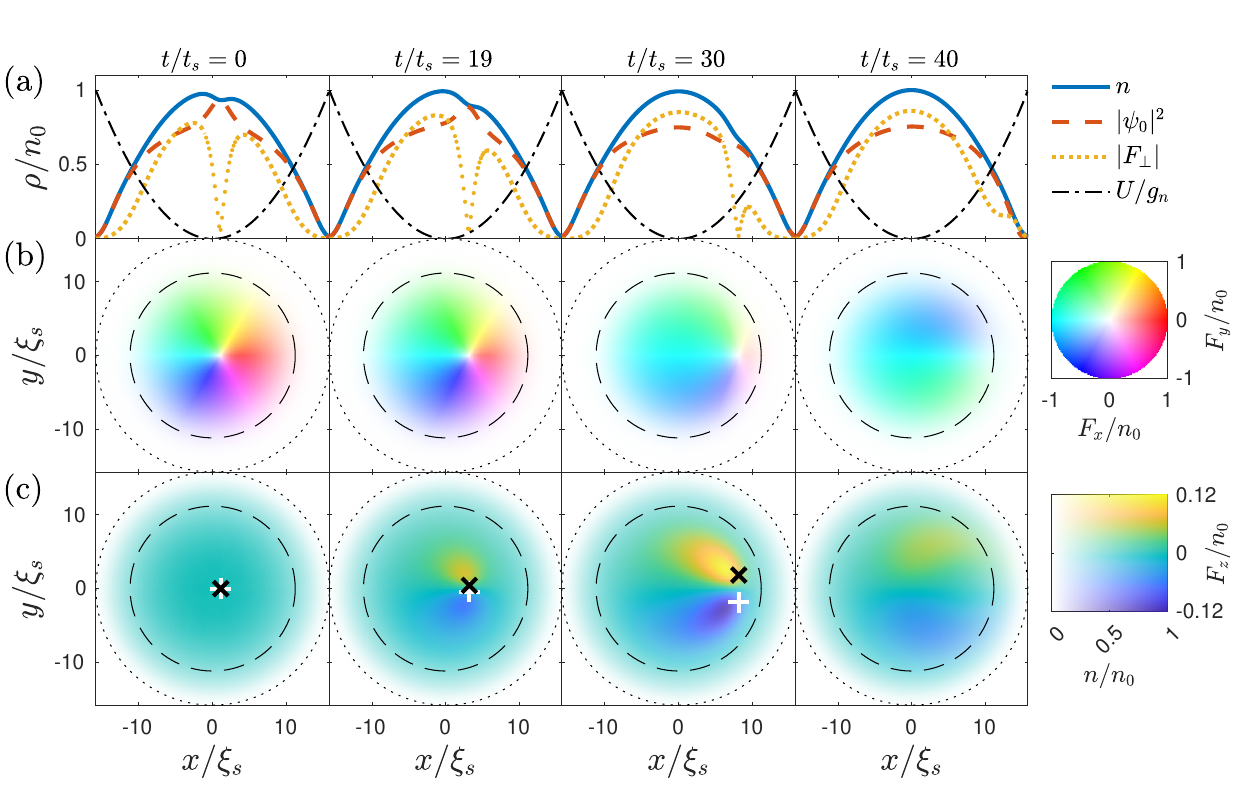}
  \caption{ \label{fig:harmonic_trap_1} Dynamics of a PCV in a harmonic trap for $q=0.5q_0$ and $\omega=0.2q_0$. (a) Cross-sectional densities $\rho=n(x,0)$ (solid blue lines), $\rho=|\psi_0(x,0)|^2$ (dashed red lines) and $\rho=|F_\perp(x,0)|$ (yellow dots). The trap $\rho=U(x,0)/g_n$, Eq.~\eqref{Uharm}, is shown for comparison (dot-dashed black lines). (b) Transverse and (c) axial spin densities. The vortex core moves radially and also ``stretches'', which is associated with the formation of a dipole of $F_z$ magnetization.  Dashed (dotted) circles in (b) and (c) are $r_{F_\perp}$ ($r_\mathrm{TF}$). White pluses (black crosses) in (c) mark centre of circulation of $\psi_1$ ($\psi_{-1}$) vortices.}
\end{figure}

\begin{figure}[!t]
  \includegraphics[trim=0cm 6cm 0cm 7cm,clip=true,width=\columnwidth]{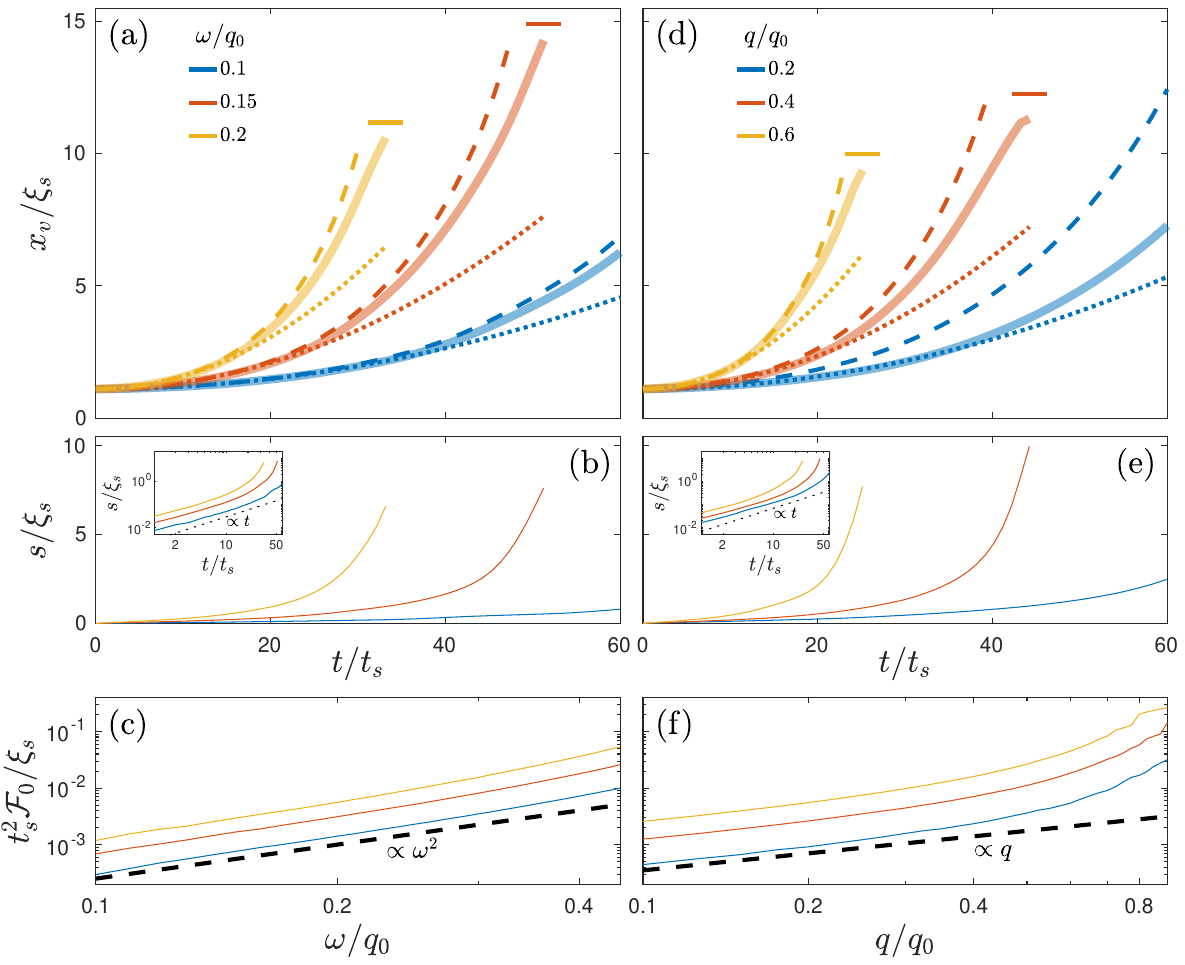}
  \caption{ \label{fig:harmonic_trap_2} (a) Evolution of PCV position for varying $\omega$ with $q=0.5q_0$. Solid lines are numerical results, dashed lines are obtained from numerically integrating the right-hand side of Eq.~\eqref{fapprox1} and dotted lines are $x_v(0)+\frac{1}{2}\mathcal{F}_0 t^2$. Short horizontal lines mark $r_{F_\perp}$. (b) The corresponding magnitude of PCV stretching. Inset: same results on a log-log scale showing regime where $s$ grows close to linearly in time and hence $\dot{s}$ is approximately constant (dotted line is $\propto t$). (c) The initial forcing $\mathcal{F}_0$ obtained from the fits in (a) increases $\propto \omega^2$, consistent with Eq.~\eqref{fapprox2}. (d)-(f) As for (a)-(c) but for varying $q$ with $\omega=0.2q_0$. The initial forcing increases $\propto q$ for small $q/q_0$, consistent with Eq.~\eqref{fapprox2}.}
\end{figure}

\section{PCV dynamics in a harmonic trap}\label{sec:harmonic}
We firstly explore PCV dynamics in a harmonic potential,
\begin{equation}\label{Uharm}
    U(\mathbf{r})=\frac{1}{2}M\omega^2 r^2.
\end{equation}
A PCV exactly at the centre of a harmonic potential is in unstable equilibrium. According to Eq.~\eqref{fapprox2}, an off-centre PCV will tend to move in the direction of increasing potential, i.e.\ to the condensate boundary. To test this, a positive, singly charged PCV is imprinted off-centre at $\mathbf{x}_v(0)=(\xi_s,0)$ with wavefunction (see Eq.~\eqref{pcvstate})
\begin{equation}\label{PCVimprint}
    \Psi(\mathbf{r})=\sqrt{\frac{n_\mathrm{TF}(\mathbf{r})}{2}}\left(\begin{array}{c}\sin\beta e^{-i\phi(\mathbf{r}-\mathbf{x}_v(0))}\\\sqrt{2}\cos\beta \\\sin\beta e^{i\phi(\mathbf{r}-\mathbf{x}_v(0))}\end{array}\right).
\end{equation}
We then evolve the system in imaginary time\footnote{The imaginary-time equations are Eq.~\eqref{eq:time_evolution_psi} with $dt\mapsto -idt$ and including a term $-\mu\psi_m$ on the right-hand side, with $\mu=(g_n+g_s)n_0+q/2$ the chemical potential~\cite{Kawaguchi2012R}.} for $-10t_s\leq t<0$, allowing the density profile to stabilise and the PCV core to form. The component phase profiles are reset to the values in Eq.~\eqref{PCVimprint} throughout the imaginary-time evolution to ensure the PCV retains its position. The state $\Psi$ is then evolved using Eq.~\eqref{eq:time_evolution_psi}. Here and throughout the paper we simulate results on a $512\times512$ grid using fourth-order Runge-Kutta integration with kinetic energy operator evaluated to spectral accuracy. We use a system size of $100\xi_s\times 100\xi_s$ with $n_0=10^4\xi_s^{-2}$.

The dynamics of the PCV in the harmonic potential is shown in Fig.~\ref{fig:harmonic_trap_1}. The PCV accelerates radially outward, as predicted by Eq.~\eqref{fapprox2}. Associated with this, the vortex cores in the $\pm 1$ spin components separate orthogonal to the PCV motion. As this occurs it becomes energetically favourable for the core of the $m=+1$ vortex to be partially filled by the $m=-1$ component and vice versa, hence the $F_z$ magnetization develops a dipole structure, see Fig.~\ref{fig:harmonic_trap_1}(c). The PCV position and stretching are shown in Fig~\ref{fig:harmonic_trap_2} for varying trap frequencies and quadratic Zeeman energies. The PCV acceleration is larger for larger trap frequency and larger quadratic Zeeman energy. Integrating the right-hand side of Eq.~\eqref{fapprox1} using numerical data for $\mathbf{s}$ allows us to test the validity of Eq.~\eqref{fapprox1} without evaluating Eq.~\eqref{sapprox}. We find good agreement with the exact numerical results for $q\gtrsim 0.3q_0$, particularly for early times. We evaluate $|\psi_0(\mathbf{x}_v)|^2$ in Eq.~\eqref{fapprox1} using Eq.~\eqref{psi0TF}, which gives a good estimate for the background component density. For later times, the stretch coordinate becomes large (see Fig~\ref{fig:harmonic_trap_2}(b),(e)) and hence we expect the approximation Eq.~\eqref{fapprox1} to become less valid. The PCV sheds spin excitations as it crosses $r_{F_\perp}$ and leaves the condensate, see Fig.~\ref{fig:harmonic_trap_1}. Note in some cases the PCV oscillates near $r_{F_\perp}$ before leaving the condensate, likely due to interactions with the emitted spin excitations.

For early times $\mathbf{s}$ grows close to linearly in time and hence $\dot{\mathbf{s}}$ is approximately constant (see insets to Fig.~\ref{fig:harmonic_trap_2}(b),(e)). We may then approximate $x_v$ by
\begin{equation}
    x_v(t)\approx x_v(0)+\frac{1}{2}\mathcal{F}_0 t^2\hspace{0.5cm}\text{(early times)},
\end{equation}
see Fig~\ref{fig:harmonic_trap_2}(a),(d). Here $\mathcal{F}_0$ is a constant forcing, obtained by fitting $x_v-x_v(0)$ to $(\mathcal{F}_0/2) t^2$ during the first half of the PCV's dynamics. We find $\mathcal{F}_0$ increases approximately quadratically with trap frequency, see Fig.~\ref{fig:harmonic_trap_2}(c), and linearly with quadratic Zeeman energy for $q/q_0\lesssim 0.4$, see Fig.~\ref{fig:harmonic_trap_2}(f). This is consistent with the qualitative approximation Eq.~\eqref{fapprox2} after expanding to linear oder in $q$. The increase in forcing with trap frequency follows directly from the scalar analysis, since the component vortices will circulate (stretch) more rapidly for tighter traps. We find $\mathcal{F}_0$ extrapolates to zero as $q\rightarrow 0$, which deviates from the prediction Eq.~\eqref{fapprox2}. We also find that Eq.~\eqref{fapprox1} overestimates the PCV acceleration for small $q/q_0$, see Fig.~\ref{fig:harmonic_trap_2}(d). These deviations may be caused by local spin rotations out of the transverse plane, which fundamentally alter the nature of the defects~\cite{Ho1998a,Lovegrove2012a,Lovegrove2016a,Williamson2017a}. An exploration of the stability of an off-centre PCV at $q=0$ would reveal if the dynamics indeed does go to zero, or if it just slows down substantially.

\section{PCV oscillations around a density minimum}\label{sec:osc}
\begin{figure}
  \includegraphics[trim=0cm 8.5cm 0cm 9cm,clip=true,width=\columnwidth]{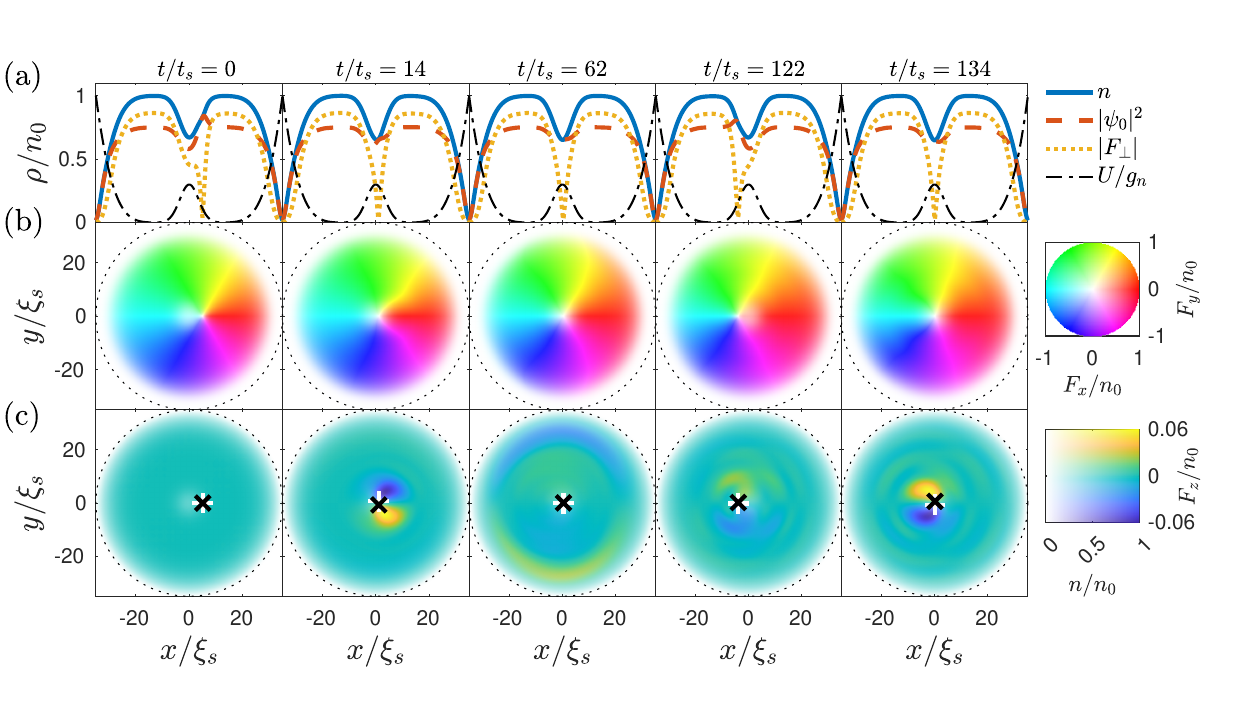}
  \caption{\label{fig:mexican_hat_1} Dynamics of a PCV for the potential Eq.~\eqref{Uinvert}, with $q=0.5q_0$, $r_0=5\xi_s$, $U_0=0.3\mu_\mathrm{TF}$ and $r_\mathrm{box}=35\xi_s$. (a) Cross-sectional densities $\rho=n(x,0)$ (solid blue lines), $\rho=|\psi_0(x,0)|^2$ (dashed red lines) and $\rho=|F_\perp(x,0)|$ (yellow dots). The potential $\rho=U(x,0)/g_n$, Eq.~\eqref{Uinvert}, is shown for comparison (dot-dashed black lines). (b) Transverse and (c) axial spin densities. The PCV moves down density gradients to the trap centre. Axial spin excitations are emitted, radiate to the condensate boundary, and then reflect back to displace the PCV again. Dotted circles in (b) and (c) are $r_\mathrm{box}$. White pluses (black crosses) in (c) mark centre of circulation of $\psi_1$ ($\psi_{-1}$) vortices.}
\end{figure}

The tendency for a PCV to move down density gradients means that a PCV will be stable at a local potential maximum. Such a potential can therefore pin a PCV in principle. To explore the dynamics of this process, we examine the dynamics of a PCV in a potential
\begin{equation}\label{Uinvert}
U(\mathbf{r}) = U_\mathrm{box}(\mathbf{r}) + U_0\exp\left(-\frac{r^2}{r_0^2}\right),
\end{equation}
where $U_0>0$ and $r_0$ characterise the height and width respectively of the local potential maximum. The approximately box-shaped trap
$U_\mathrm{box}(\mathbf{r})=\mu_\mathrm{TF}\coth^{10}(1)\tanh^{10}(r/r_\mathrm{box})$
characterised by size $r_\mathrm{box}\gg r_0$ confines the condensate~\cite{gaunt2013,navon2021}. The $\coth^{10}(1)$ prefactor is included so that $r_\mathrm{TF}\approx r_\mathrm{box}$ for $r_\mathrm{box}\gg r_0$. The potential Eq.~\eqref{Uinvert} results in a local maximum in condensate density at the trap centre, see Fig.~\ref{fig:mexican_hat_1}(a). A positive, singly charged PCV is imprinted at $\mathbf{x}_v(0)=(5\xi_s,0)$ using Eq.~\eqref{PCVimprint}. Further numerical details are the same as for the harmonic trap, see Sec.~\ref{sec:harmonic}. The potential Eq.~\eqref{Uinvert} increases radially inward for $r\lesssim r_\mathrm{box}$, and hence the PCV initially moves radially inward, see Fig.~\ref{fig:mexican_hat_1}.

\begin{figure}
  \includegraphics[trim=0cm 4cm 0cm 4cm,clip=true,width=\columnwidth]{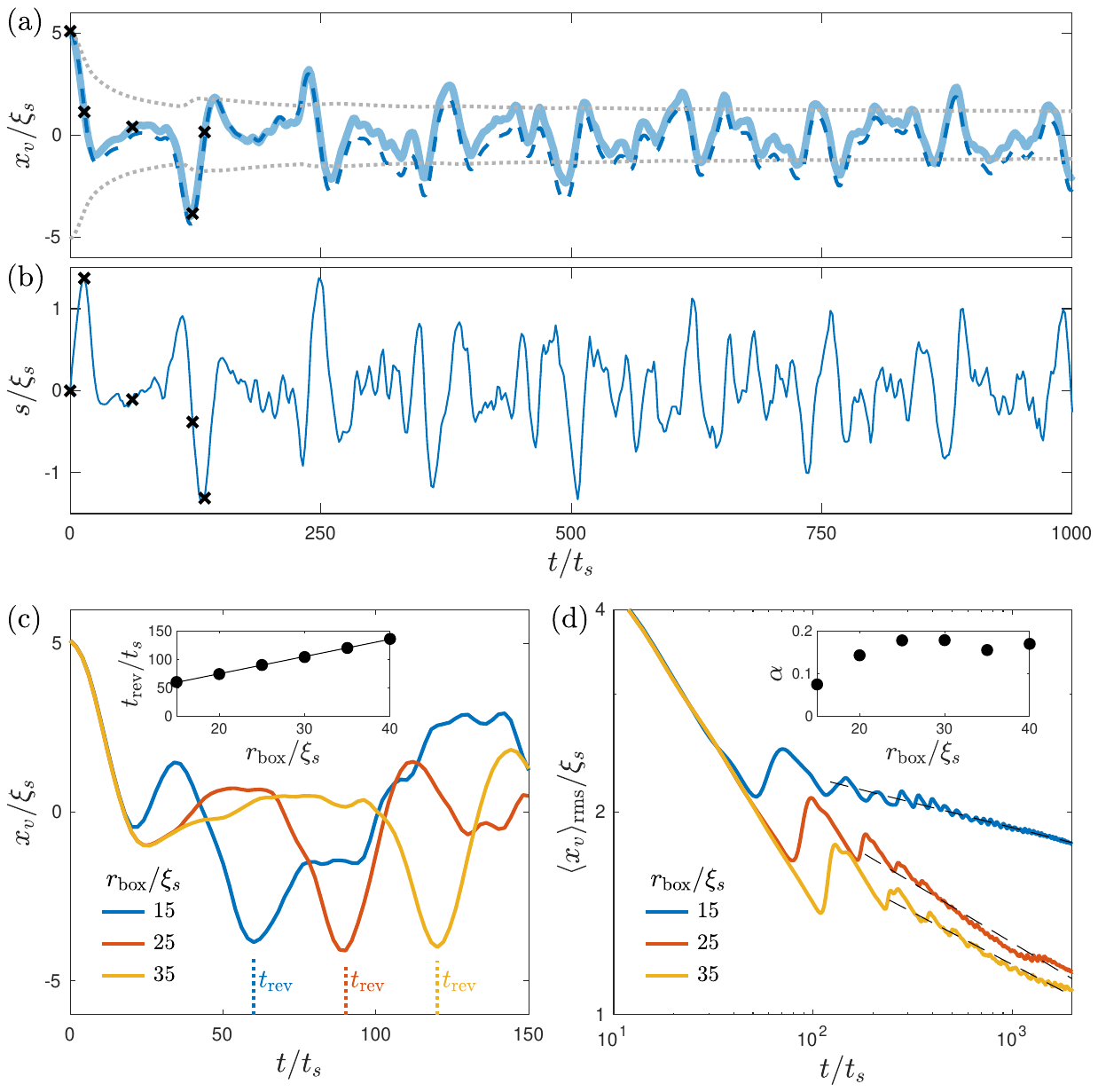}
  \caption{ \label{fig:mexican_hat_2} Evolution of (a) the PCV position and (b) PCV stretching, with parameters as in Fig.~\ref{fig:mexican_hat_1}. The PCV position oscillates around the trap centre. Solid lines are the numerical results, dashed-blue line in (a) is obtained from numerically integrating the right-hand side of Eq.~\eqref{fapprox1}, crosses correspond to times in Fig.~\ref{fig:mexican_hat_1}. The PCV stretching peaks when the PCV has a large velocity. (c) PCV dynamics showing oscillations with increasing box size. The magnitude of $x_v$ revives at a time $t_\mathrm{rev}$ (vertical dotted lines), which increases with box size.  Inset: $t_\mathrm{rev}$ (circles) is fitted by $t_\mathrm{rev}/t_s\approx 3r_\mathrm{box}/\xi_s+14$ (line), consistent with a constant axial spin-wave propagation speed. (d) The rms amplitude $\langle x_v\rangle_\mathrm{rms}$ (Eq.~\eqref{xrms}) decays as $t^{-\alpha}$ at long times (black-dashed lines) with exponent $\alpha\sim 0.1$-$0.2$ (inset). Dotted-gray lines in (a) are $\pm \langle x_v\rangle_\mathrm{rms}/\xi_s$. All results are for $q=0.5q_0$, $r_0=5\xi_s$ and $U_0=0.3\mu_\mathrm{TF}$.}
\end{figure}

Axial spin excitations are emitted as the PCV moves toward the trap centre. These excitations propagate out to and then reflect off the condensate boundary, see Fig.~\ref{fig:mexican_hat_1}(c). The PCV stretching increases as the PCV velocity increases. Both the PCV position and stretching plateau near the trap centre. Rather than remaining here, however, the PCV position undergoes periodic kicks to a displacement on the order of the initial PCV position, see Fig.~\ref{fig:mexican_hat_2}\ (a),\ (c). This revival in PCV displacement coincides with axial spin excitations localising close to the trap centre after reflecting off the condensate boundary. We interpret this as a re-excitation of the PCV motion due to absorption of axial spin excitations, which results in a rephasing of the transverse spin. Evidence for this is seen by increasing $r_\mathrm{box}$, which increases the time for spin excitations to propagate to and from the condensate boundary, and hence the time $t_\mathrm{rev}$ for the PCV displacement to revive, see Fig.~\ref{fig:mexican_hat_2}(c). We find that $t_\mathrm{rev}$ is well described by a linear fit,
\begin{equation}\label{trevfit}
    t_\mathrm{rev}\approx \frac{2r_\mathrm{box}}{v_\mathrm{fit}}+t_\mathrm{offset},
\end{equation}
with $v_\mathrm{fit}\approx 0.7\xi_s/t_s$ and $t_\mathrm{offset}\approx 14t_s$ fitting parameters, see inset to Fig.~\ref{fig:mexican_hat_2}(c). The accuracy of the fit is consistent with a constant spin-wave propagation speed of $v_\mathrm{fit}=2r_\mathrm{box}/(t_\mathrm{rev}-t_\mathrm{offset})$. Furthermore, $v_\mathrm{fit}$ is close to the speed of sound of the gapless spin mode in a uniform system, which for $q=0.5q_0$ is $v_s=0.5\xi_s/t_s$~\cite{Uchino2010}. The small offset $t_\mathrm{offset}$ in Eq.~\eqref{trevfit} may account for the time for spin waves to interact with the PCV and the condensate boundary. The stretch coordinate of the PCV oscillates out of phase with the PCV displacement, acquiring a maximum value when the PCV is moving most rapidly, see Fig.~\ref{fig:mexican_hat_2}(b).

The PCV continues to oscillate around the trap centre for long times $t\gg t_\mathrm{rev}$, see Fig.~\ref{fig:mexican_hat_2}(a). There is a weak decay in the oscillation amplitude with time, which we characterise by plotting the rms amplitude
\begin{equation}\label{xrms}
    \langle x_v\rangle_\mathrm{rms}(t)=\sqrt{\frac{1}{t} \int_0^t d\tau\,x_v(\tau)^2}.
\end{equation}
We find $\langle x_v\rangle_\mathrm{rms}$ can be fitted by a power law $At^{-\alpha}$ for long times $t>2t_\mathrm{rev}$, see Fig.~\ref{fig:mexican_hat_2}(d). For comparison, $\langle x_v\rangle_\mathrm{rms}(t)$ for a perfectly sinusoidal oscillation would approach a constant for large $t$. For an initial PCV position $x_v(0)=5\xi_s$ as in Fig.~\ref{fig:mexican_hat_2} we find $\alpha\sim 0.1$-$0.2$, see inset to Fig.~\ref{fig:mexican_hat_2}(d). The value of $\alpha$ is sensitive to the initial PCV position, tending to decrease for an initial position closer to the trap centre. We expect the PCV oscillations will damp out after the axial spin waves thermalize. The weak decay in Fig.~\ref{fig:mexican_hat_2}(d) suggests this thermalization happens slowly, consistent with prior work on phase ordering~\cite{williamson2019}.

\begin{figure}
  \includegraphics[trim=0cm 9.5cm 0cm 10cm,clip=true,width=\columnwidth]{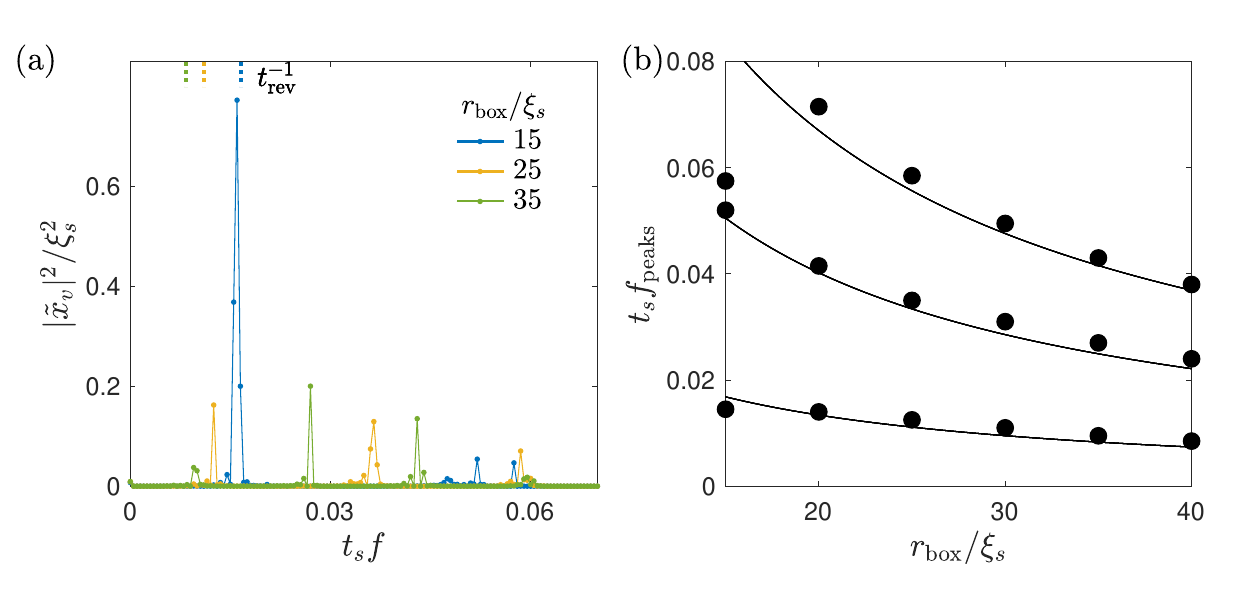}
  \caption{ \label{fig:powerspectra} (a) The power spectra of $x_v$ (Eq.~\eqref{powerspec}) exhibit narrow, regularly spaced peaks corresponding to the oscillations in Fig.~\ref{fig:mexican_hat_2}. The lowest frequency of these corresponds to $t_\mathrm{rev}^{-1}$ (vertical dotted lines). (b) The frequency of the peaks in $|\tilde{x}_v|^2$ (circles) decrease with increasing box size. Lines are $t_\mathrm{rev}^{-1}$, $3t_\mathrm{rev}^{-1}$ and $5t_\mathrm{rev}^{-1}$, with $t_\mathrm{rev}$ given by Eq.~\eqref{trevfit}. Only peaks with $|\tilde{x}_v|^2\ge 0.02\xi_s^2$ and spaced $\ge 0.004t_s^{-1}$ from another peak are displayed. All results are for $q=0.5q_0$, $r_0=5\xi_s$ and $U_0=0.3\mu_\mathrm{TF}$.}
\end{figure}

Finally, we compute the power spectrum of $x_v$, defined as
\begin{equation}\label{powerspec}
    |\tilde{x}(f)|^2=\left|2T^{-1}\operatorname{Re}\int_0^T x_v(t) e^{-i2\pi f t}\,dt\right|^2,
\end{equation}
with $T$ the evolution time. These are shown in Fig.~\ref{fig:powerspectra}(a). The power spectra exhibit narrow peaks at frequencies $f_\mathrm{peaks}$ with $f_\mathrm{peaks}$ decreasing with increasing box size, see Fig.~\ref{fig:powerspectra}(b). The lowest frequency of these corresponds to $t_\mathrm{rev}^{-1}$, consistent with Fig.~\ref{fig:mexican_hat_2}(c). In addition there are pronounced peaks at $3t_\mathrm{rev}^{-1}$ and $5t_\mathrm{rev}^{-1}$ (with additional smaller peaks at higher multiples of $t_\mathrm{rev}^{-1}$). The generation of these frequency harmonics may be due to non-linear effects within the PCV or due to non-linear interactions between the axial spin waves.

Qualitatively similar vortex oscillations are obtained when the box trap is replaced by a softer-walled trap such as a harmonic trap. The quantitative analysis, however, is more complicated, as the propagation speed of axial spin waves becomes spatially dependent and collective excitations of the condensate (breathing and higher-order modes) may be excited.

\section{Energy considerations}\label{sec:energy}
The azimuthal motion of a vortex in a harmonically trapped scalar condensate ensures the kinetic energy contained in the flow field is conserved~\cite{groszek2018}. In contrast, the kinetic energy in the flow field of a PCV is not conserved during its radial motion. In this section we explore the energy exchanges that occur during the PCV dynamics.

\begin{figure}
 \includegraphics[trim=0cm 9.5cm 0cm 10cm,clip=true,width=\columnwidth]{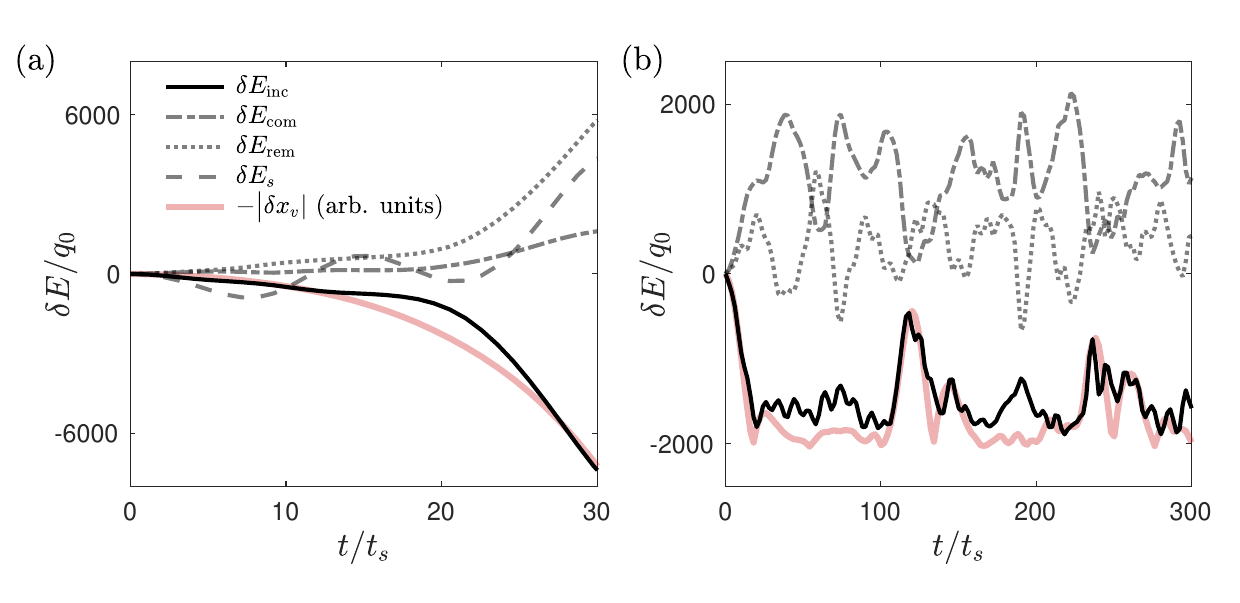}
\caption{\label{energies} Energy exchanges during the PCV dynamics for both trapping potentials Eq.~\eqref{Uharm} and Eq.~\eqref{Uinvert}, with $\delta E_\mu(t)\equiv E_\mu(t)-E_\mu(0)$ and $\delta x_v(t)\equiv |x_v(t)|-|x_v(0)|$. The incompressible energy $E_\mathrm{inc}$ decreases on average as the PCV moves to regions of lower density, with $\delta E_\mathrm{inc}\sim -|\delta x_v|$. (a) The harmonic trap, Eq.~\eqref{Uharm}. Here $E_\mathrm{inc}$ is primarily converted into $E_\mathrm{rem}$ (predominantly $E_s$), with a small amount converted into $E_\mathrm{com}$ (parameters as in Fig.~\ref{fig:harmonic_trap_1}). (b) The potential with a local maximum at the centre, Eq.~\eqref{Uinvert}. Here $E_\mathrm{inc}$ is primarily converted into $E_\mathrm{com}$, consistent with the PCV radiating axial spin waves (parameters as in Fig.~\ref{fig:mexican_hat_1}).}
\end{figure}

The contribution to the kinetic energy from the PCV flow field is
\begin{equation}
    E_\mathrm{inc}=\frac{\hbar^2}{2M}\sum_{m=-1,1} \int \dd^2\br\,|\mathbf{v}_m^\mathrm{inc}|^2,
\end{equation}
where $\mathbf{v}_m^\mathrm{inc}$ is the incompressible (divergence-free) component of the density-weighted velocity field $\mathbf{v}_m=|\psi_m|^{-1}\operatorname{Im}\left(\psi_m^*\nabla\psi_m\right)$~\cite{nore1997,Bradley2012a}. Note $\mathbf{v}_0=0$ due to our choice of initial condition. Figure~\ref{energies} shows that $E_\mathrm{inc}$ decreases as the PCV centre --- the region of highest fluid velocity --- moves to lower density. The fluid flow also has a compressible component
\begin{equation}
    E_\mathrm{com}=\frac{\hbar^2}{2M}\sum_{m=-1,1} \int \dd^2\br\,|\mathbf{v}_m^{\text{com}}|^2,
\end{equation}
with $\mathbf{v}_m^{\text{com}}$ the compressible (curl-free) component of $\mathbf{v}_m$. We denote the remaining energy terms by
\begin{equation}
    E_\mathrm{rem}=H-E_\mathrm{inc}-E_\mathrm{com}.
\end{equation}

In a harmonic trap, Eq.~\eqref{Uharm}, $E_\mathrm{inc}$ is monotonically converted predominantly into $E_\mathrm{rem}$, see Fig.~\ref{energies}(a). The predominant contribution to $E_{\text{rem}}$ is the spin energy
\begin{equation}
E_s=\int\dd^2\br\,\left[q\left(|\psi_1|^2+|\psi_{-1}|^2\right)+\frac{g_s}{2}|\mathbf F(\br)|^2\right],    
\end{equation}
see Fig.~\ref{energies}\,(a), consistent with this energy arising due to PCV stretching.
We suspect the oscillation in $E_s$ is due to collective modes in the system coupling both spin and density degrees of freedom.

In the potential Eq.~\eqref{Uinvert}, $E_\mathrm{inc}$ oscillates with the oscillating PCV position, see Fig.~\ref{energies}(b). Here the times at which the PCV is close to the trap centre coincide with an average increase in $E_\mathrm{com}$ rather than $E_\mathrm{rem}$. This is consistent with the energy arising from radiated axial spin waves. Peaks of $E_\mathrm{inc}$ occur when the PCV is at maximum displacement from the trap centre, and this coincides with troughs of $E_\mathrm{com}$.

\section{Conclusion}\label{sec:conclusion}
We have demonstrated that a PCV moves down density gradients in an inhomogeneous condensate. Our results are well-described by a simplified model, Eq.~\eqref{fapprox1}, adapted from a model of vortex dynamics in scalar condensates. We have considered spin interactions $|g_s|\ll g_n$, which we expect to be representative of the behaviour of $^{87}$Rb condensates. Experiments using $^{87}$Rb have observed spin dynamics on the order of $10^2\,t_s$~\cite{prufer2018,prufer2022}, which is sufficient to observe the dynamics of a PCV in moderately confined harmonic traps. A $^7$Li condensate also exhibits ferromagnetic spin interactions but with a much larger interaction strength $|g_s|$, enabling observation of spin dynamics on the order of $10^3\, t_s$~\cite{huh2020,huh2023}. However, in $^{7}$Li the spin and density interaction energies are comparable ($g_n/|g_s|\approx 2$)~\cite{huh2020}, which will affect the background condensate profile and quasiparticle excitations in the system. This will likely modify the acceleration of a PCV and affect the oscillations observed in Sec.~\ref{sec:osc}. The PCV dynamics and axial spin excitations could be observed in experiments using \emph{in situ} imaging~\cite{Higbie2005a,Sadler2006a,Seo2015a,Seo2016a}.

Our results are robust to weak external damping $\gamma\lesssim 10^{-3}$, which is included in the dynamics via the replacement $dt\rightarrow (1-i\gamma dt)$~\cite{choi1998,rooney2010}. Larger values of damping $10^{-2}\lesssim \gamma\lesssim 1$ tend to reduce the PCV stretching and hence decelerate the PCV motion, whereas very strong damping $\gamma>1$ increases the PCV acceleration. In addition, the oscillations in Fig.~\ref{fig:mexican_hat_2}(a),(b) decrease with $\gamma$ for $\gamma\gtrsim 10^{-2}$ and disappear completely for $\gamma\gtrsim 0.1$. A more thorough exploration of the damped and finite-temperature dynamics of PCVs is an interesting prospect for future work.

Exploring in detail the structure of the PCV core throughout the dynamics and how this interacts with other excitations in the condensate would be an interesting area for future work, particularly considering the vortex stretching is prone to intrinsic damping~\cite{williamson2021}. We expect a PCV moving at a sufficiently large velocity to destabilise, while preserving the total topology~\cite{Turner2009}; exploring this would be an interesting area for future work. Exploring the effect of a net axial spin magnetization would also be interesting, which we expect will allow the vortex dynamics to be tuned between PCV-like and scalar-like dynamics~\cite{williamson2021}. Trap engineering can be used to create complex condensate density profiles~\cite{gauthier2016} and may allow steering of the PCV motion, resulting in the possibility of rich but controllable vortex trajectories. Finally, exploring the interplay between inhomogeneity and additional PCVs would be interesting, particularly in relation to nonlinear phenomena such as turbulence and the dynamics of phase transitions.

\section*{Acknowledgements}
We thank Andrew Groszek, Blair Blakie and Matthew Reeves for valuable discussions.

\paragraph{Funding information}
This research was supported by the Australian Research Council Centre of Excellence for Engineered Quantum Systems (EQUS, CE170100009) and the Australian government Department of Industry, Science, and Resources via the Australia-India Strategic Research Fund (AIRXIV000025).

\end{document}